\documentclass[prb, superscriptaddress,reprint,amsmath,longbibliography]{revtex4-2}
\usepackage{graphicx}
\usepackage{amssymb}
\usepackage{physics}

\usepackage{refcount}
\usepackage{comment}
\usepackage{caption}
\usepackage{ragged2e}
\usepackage[utf8]{inputenc}
\usepackage{float}
\addtocontents{toc}

\usepackage[format=plain,justification=raggedright]{caption}

\usepackage[
    colorlinks=true,
    linkcolor=blue,
    citecolor=blue,
    urlcolor=blue,
    filecolor=blue,
    pdfborder={0 0 0}
]{hyperref}
\begin{document}

% Save the normal TOC-writing command
\let\savedaddcontentsline\addcontentsline

% Do NOT write the main-paper sections into the TOC
\renewcommand{\addcontentsline}[3]{}
\title{Switchable giant room-temperature nonlinear Hall effect in Bilayer Graphene}

\author{Margherita Melegari}
\email{margherita.melegari@unige.ch}
\affiliation{Department of Quantum Matter Physics, University of Geneva, Quai Ernest-Ansermet 24, 1211 Geneva, Switzerland}
\affiliation{Department of Applied Physics, University of Geneva, 24 Quai Ernest Ansermet, Geneva, CH-1211 Switzerland}

\author{Ignacio Gutiérrez-Lezama}
\affiliation{Department of Quantum Matter Physics, University of Geneva, Quai Ernest-Ansermet 24, 1211 Geneva, Switzerland}
\affiliation{Department of Applied Physics, University of Geneva, 24 Quai Ernest Ansermet, Geneva, CH-1211 Switzerland}

\author{Justin C. W. Song}
\affiliation{Division of Physics and Applied Physics, School of Physical and Mathematical Sciences, Nanyang Technological University, Singapore 637371, Singapore}
\author{Alberto F. Morpurgo}
\email{alberto.morpurgo@unige.ch}
\affiliation{Department of Quantum Matter Physics, University of Geneva, Quai Ernest-Ansermet 24, 1211 Geneva, Switzerland}
\affiliation{Department of Applied Physics, University of Geneva, 24 Quai Ernest Ansermet, Geneva, CH-1211 Switzerland}

\date{\today}
\begin{abstract}
\textbf{Utilizing quantum second-order nonlinear transport for practical junction-free devices require materials with large and tunable nonlinearites at room temperature – a current materials platform challenge. Here, we report the nonlinear Hall effect  (NLHE) in double-ionic gated bilayer graphene devices that enable unusually strong inversion breaking. We observe NLHE that are readily switchable (on, off, and sign reversed) with second order nonlinear susceptibilities $\chi^{(2)}_{yxx}$ that reaches giant room-temperature values of 3 10$^{-3} \mu$m S/V, comparable to values commonly observed at low temperature in WTe$_2$ or in graphene-based moiré superlattices, and three-to-four orders of magnitude larger than values reported in material systems recently employed in search of a room-temperature NLHE. Our devices produce corresponding THz voltage responsivities $\simeq 4 \, 10^{4}$ V/W, comparable to commercially available Schottky diodes. These are orders of magnitude better than for previously reported room-temperature NLHE devices rendering double-ionic gated bilayer graphene a choice platform for junction-free nonlinear technology.} 
\end{abstract}

\maketitle

\section{Introduction}
Nonlinear phenomena in time-reversal-invariant conductors attract interest as they enable the investigation of the electronic structure of quantum materials, including different aspects of topology \cite{sodemann.2015,du.2018,facio.2018,wang.2019,zhang.2018,ma.2019,matsyshyn.2019,rostami.2020,tu.2020,sinha.2022,yin.2022,gao.2023,wang.2023,qin.2024,mercaldo.2025,suarez.2025,yu.2025,zhu.2025}, scattering mechanisms \cite{du.2019,du.2021b,xiao.2019,nandy.2019,isobe.2020,he.2022,tiwari.2021,shvetsov.2019,ma.2023,makushko.2024,suarez.2025}, and symmetry \cite{sodemann.2015,nandy.2019,du.2021,ideue.2021,ortix.2021,itahashi.2022,duan.2023,suarez.2024,suarez.2025,zhu.2025}. In contrast to trivial nonlinearities originating from Joule heating and other thermally driven mechanisms \cite{cao.2011, el_filali.2022}--which occur irrespective of the system symmetry-- second-order nonlinear phenomena in time reversal invariant conductors require inversion symmetry to be broken \cite{sodemann.2015,nandy.2019,du.2021b,xiao.2019,ortix.2021,bandyopadhyay.2024,yu.2025}. This dependence on the absence of inversion symmetry makes nonlinear phenomena sensitive to electronic properties not directly accessible through linear response measurements. 

The same phenomena also provide the working principle of devices that aim at rectifying high-frequency electrical signals \cite{isobe.2020,zhang.2021,kumar.2024,makushko.2024}, potentially useful in sensing, transduction, and energy harvesting \cite{shi.2023,onishi.2024}. In this context, the nonlinear Hall effect (NLHE) is the second-order nonlinear transport phenomenon that has attracted most attention \cite{du.2021,ortix.2021,bandyopadhyay.2024,ideue.2021,suarez.2025}.  When an alternating current flows at frequency $\omega$, the NLHE generates --in the absence of any applied magnetic field-- a transverse voltage  that contains a rectified and a second-harmonic contribution $V_{xy}^{(2\omega)}$ at $2\omega$, whose  amplitude scales quadratically with current \cite{sodemann.2015,ma.2019,kang.2019,shvetsov.2019,matsyshyn.2019}. For practical applications, materials are needed that exhibit the largest magnitude of the NLHE at room temperature. This requirement poses a challenge, because in most systems studied the NLHE signal becomes immeasurably small at room temperature \cite{ma.2019,kang.2019,tiwari.2021,zhang.2018,zeng.2019,he.2022,sinha.2022,huang.2023}. A handful of compounds have been identified that produce measurable NLHE at room temperature (TaIrTe$_4$, BaMnSb$_2$, elemental Bismuth) \cite{kumar.2021, min.2023,makushko.2024}, but  the magnitude of the measured NLHE signal is three-to-four orders of magnitude smaller than what is observed at low temperature in WTe$_2$ or graphene-based moiré superlattices.  \cite{zhang.2018,ma.2019,kang.2019,wang.2019,zeng.2019,tiwari.2021,he.2022,sinha.2022,huang.2023}.

Here we employ double-ionic gated Bernal stacked bilayer graphene (BLG) devices to investigate whether a NLHE signal is induced by the application of a perpendicular electric field that breaks inversion symmetry. Double ionic gating enables the application of unprecedented perpendicular electric field \cite{domaretskiy.2022,weintrub.2022,melegari.2026} (electric fields --not displacement fields-- of more than 2 V/nm between the two graphene layers forming the BLG \cite{melegari.2026}) that allow exploring the extreme inversion symmetry breaking regime. In this regime,  the NLHE is found to emerge already at room temperature, generating a  signal that can be fully controlled: the signal reverses its sign upon inverting the polarity of the perpendicular electric field and --for each polarity-- it increases as the field strength is increased. At the largest perpendicular electric field, the NLHE signal reaches values comparable to those commonly observed in WTe$_2$ or graphene-based moiré superlattices at low temperature~\cite{zhang.2018,ma.2019,kang.2019,wang.2019,zeng.2019,tiwari.2021,he.2022,sinha.2022,huang.2023}.  We quantify the phenomenon by the nonlinear susceptibility tensor $\chi_{ijk}^{(2)}$ (defined as $J_i^{2\omega} = \chi_{ijk}^{(2)} E_j^{\omega} E_k^{\omega}$), which expresses the current $J_i^{2\omega}$ flowing at frequency $2\omega$ in terms of the applied electric field $E_j^{\omega}$ oscillating at frequency $\omega$ and find that the term responsible for the NLHE, $\chi^{(2)}_{yxx} $, reaches values of 2 10$^{-3}$ $\mu m$SV$^{-1}$, an unprecedented magnitude at room temperature \cite{zhang.2018,ma.2019,kang.2019,tiwari.2021,he.2022,huang.2023}. Our work demonstrates the ability to experimentally control the sign, and most importantly, the strength of the NLHE at room temperature, which is locked to the strength of inversion breaking, providing a clear tool for large NLHE devices.

\section{Double ionic gated graphene bilayer Hall bars}

\begin{figure}
    \centering
    \includegraphics[width=0.5\textwidth]{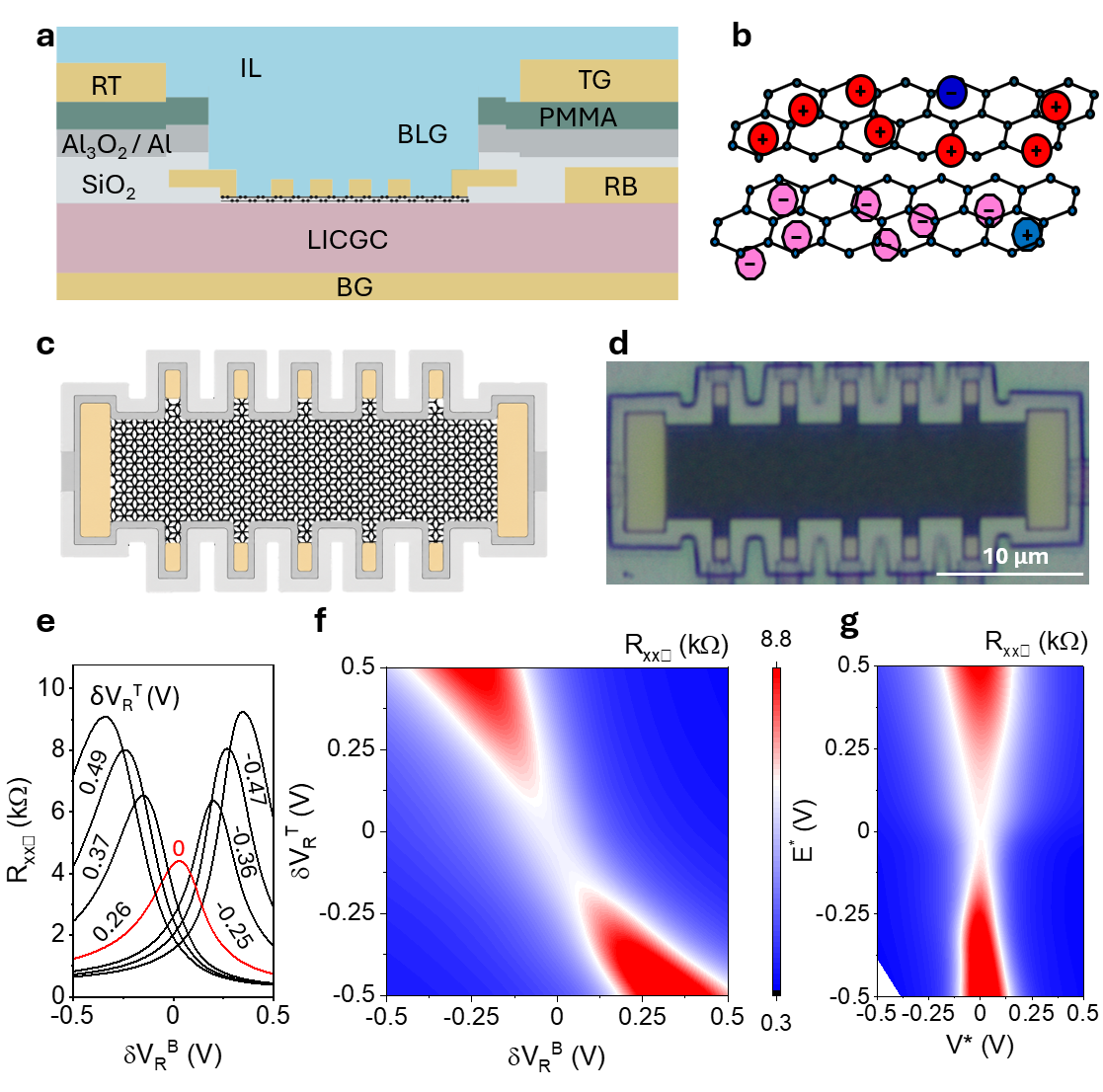}
    \caption{\justifying \textbf{Double ionic gated BLG devices.} \textbf{a}  Schematic side view of a double ionic gated Hall bar BLG device (not to scale). A BLG on a LICGC electrolyte substrate with gold electrodes is in direct contact with a top  ionic liquid (IL) electrolyte. The LICGC and the IL are electrostatically decoupled by a stack of SiO$_2$/Al/Al$_2$O$_3$/PMMA layers. The top gate (TG) and the top reference (RT) electrodes are large metallic pads (Ti/Au) deposited onto the PMMA; the back gate (BG) and bottom reference (RB) electrodes are metallic contacts (Cr/Au) deposited respectively on the back and on the top of the LICGC substrate. \textbf{b} The top and bottom layers of BLG are in direct contact with ions in the two electrolytes;  ions in the top and bottom electrolytes are different chemical species. \textbf{c} Top view of the BLG Hall bar. The grey shaded region indicates the edges of the electrostatic decoupling layer that slightly overlaps with the BLG. \textbf{d} Optical microscope image of a device prior to depositing the IL. \textbf{e} Square resistance $R_{xx\square}$  measured as a function of $\delta V_R^B$ for different values of $\delta V_R^T$ (indicated next to the curves; the red curve for which the height of the resistance peak is minimum corresponds to having zero perpendicular electric field at charge neutrality). \textbf{f} Full dependence of $R_{xx\square}$ on $\delta V_R^B$ and $\delta V_R^T$, exhibting the behavior expected for double-gated BLG. \textbf{g} Same data as in panel \textbf{f} plotted versus $V^*$ and $E^*$.}
    \label{fig:FIG1}
\end{figure}

Our experiments rely on double ionic gated Bernal-stacked bilayer graphene (BLG) devices (see Fig.\ref{fig:FIG1}a), whose detailed structure, fabrication, and characterization are discussed in the supplementary information, S1. Here we explain the aspects essential to understand the NLHE measurements. In double ionic gated BLG devices, two electrolytes --a bottom solid Li-ion conducting glass ceramic (LICGC) substrate containing a very large density of mobile Li$^+$ ions \cite{domaretskiy.2022,cao.2023,melegari.2026} and a top DEME-TFSI ionic liquid formed by positively and negatively charged mobile molecules \cite{fujimoto.2013, bisri.2017, zhang.2019, gutierrez.2021}-- are in direct contact with the BLG bottom and top layers. These electrolytes transfer  the potentials applied to two gate electrodes (see Fig.\ref{fig:FIG1}) coupled to opposite sides of the BLG. Because the density of mobile ions in the electrolytes is extremely large and the electrostatic screening length is only a few Angstroms, the potential applied to the gate drops entirely at the gate/electrolyte and the electrolyte/BLG interfaces, without any voltage drop across the bulk of the electrolytes \cite{shimotani.2006,Xia.2010,fujimoto.2013,bisri.2017,kim.2013}. The short screening length also results in extremely large geometrical capacitances $C_T$ and $C_B$ for both electrolytes,  of the order of 50 $\mu$F/cm$^2$ (the exact value depends on polarity) \cite{lee.2007,schmidt.2016,bisri.2017, philippi.2018, zhang.2019, cao.2023,melegari.2026}. 

Two reference electrodes, each coupled to one of the electrolytes, read the potentials $\delta V_R^B$ and $\delta V_R^T$ dropping at the bottom and top electrolyte/BLG interfaces. From $\delta V_R^B$ and $\delta V_R^T$ we estimate the strength of the perpendicular electric field $E_{BLG}$ between the graphene layers forming the BLG (see SI, S2), which scales approximately linearly with  $\delta V_R^B - \delta V_R^T$, (we estimate  $E_{BLG} \simeq$ 1.8 V/nm at $\delta V_R^B - \delta V_R^T = $ 1 V; see S2 in SI). Even larger potential differences can be applied and we have discussed elsewhere the linear transport properties in the regime $\delta V_R^B - \delta V_R^T > 1$ V , in which new physical phenomena become visible due to the presence of in-gap states of electrons in graphene bound to ions in the electrolytes \cite{melegari.2026}. Here, we confine the study of the NLHE to the regime for which the presence of in-gap states does not affect qualitatively the linear transport properties of BLG. 

Some specific aspects of double ionic gated devices should be emphasized. First, our study relies on measurements of linear and nonlinear transport as a function of the two gate voltages, to control independently the  breaking of inversion symmetry and the density of accumulated charge carriers. Because ionic motion in the electrolytes freezes at relatively high temperature (220 K in DEME-TFSI) below which we cannot tune the ionic gates, here we focus on room-temperature investigations of the NLHE  (few measurements performed at lower temperatures above 220 K showed no relevant differences). Second, because ions in both electrolytes are in direct contact with the BLG and scatter charge carriers,  the measured carrier mobility is relatively low (approximately 1000-2000 cm$^2$/Vs)\cite{melegari.2026}, and disorder is expected to play an important role.
Third, in the two electrolytes --and in the same electrolyte for opposite polarities-- the ions accumulated by the applied gate voltage at the interface with the BLG are different \cite{domaretskiy.2022,melegari.2026}. In the LICGC substrate mobile Li-ions carry positive charges while the compensating negative countercharges are fixed in the ceramic matrix \cite{domaretskiy.2022,cao.2023,melegari.2026}; in the ionic liquid the positive and negative molecular ions are both mobile, but have different sizes \cite{fujimoto.2013, bisri.2017, zhang.2019, gutierrez.2021}. Therefore details of the device behavior that depend on the specific ions --including the type of disorder or the exact value of the geometrical capacitance-- can differ for opposite gate voltage polarities.

We have realized three devices for the investigation of linear transport at very large perpendicular electric field, and measured non-linear transport in two of them. The behavior of the linear and nonlinear response of all devices is extremely reproducible. The data shown in the main text have been measured on one device and NLHE data from another device are shown in the supplementary information to illustrate reproducibility (the reproducibility of the linear transport response is also illustrated in the supplementary information and discussed in more detail in Ref.\cite{melegari.2026}).

\begin{figure*}[t]
    \centering
\includegraphics[width=0.8\linewidth]{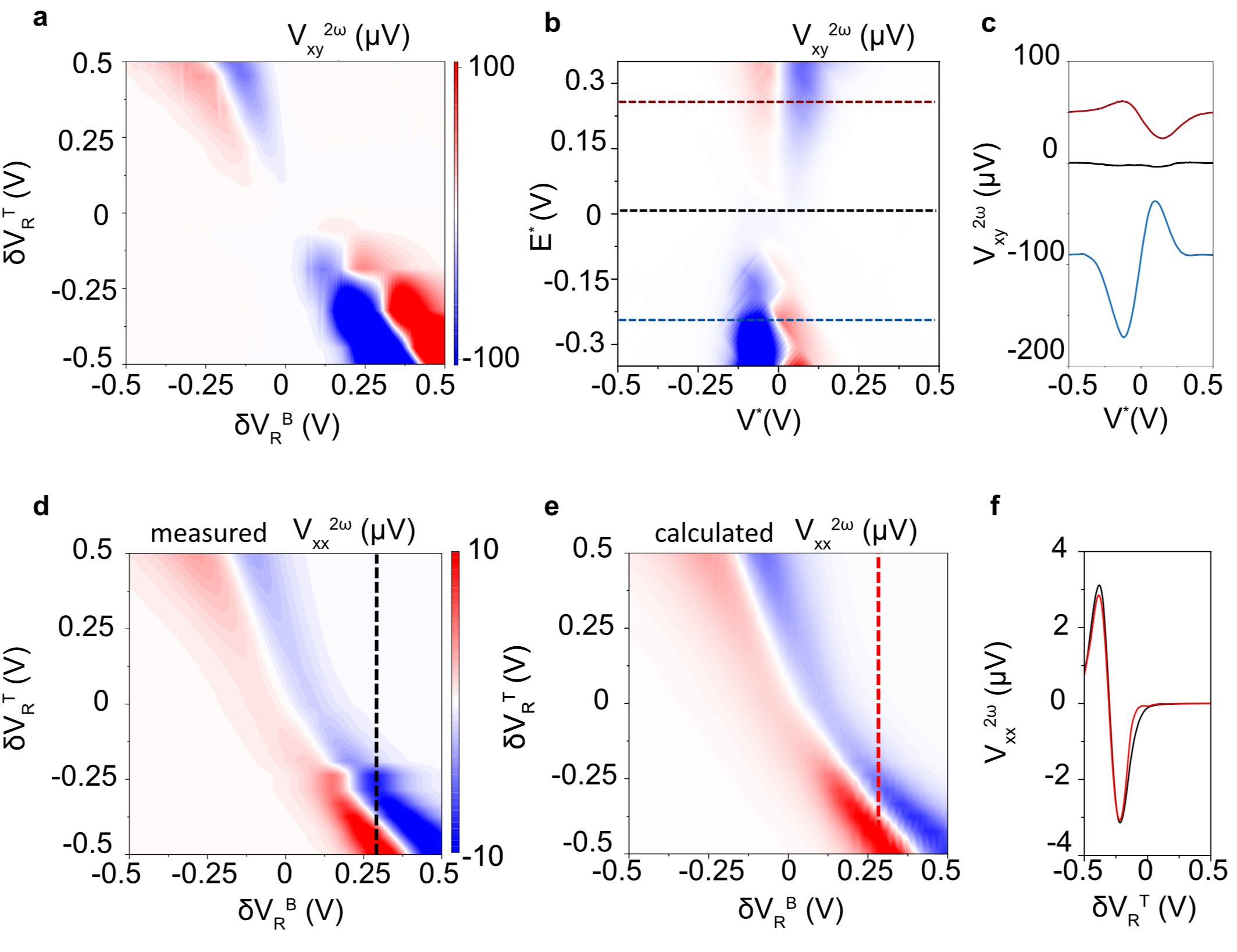}
\caption{ \justifying \textbf{Nonlinear transverse and longitudinal signals. a} Second harmonic transverse voltage $V_{xy}^{2\omega}$ measured with excitation current  $I^{\omega} = 2 \mu$A at frequency $\omega$, as a function of  $\delta V_R^B$ and $\delta V_R^T$. \textbf{b} Same data as in \textbf{a}, plotted versus $V^*$ and $E^*$.  $V_{xy}^{2\omega}$ changes sign when $V^*$ is reversed  (i.e., when the Fermi level is shifted from the valence to the conduction band) and when $E^*$ is reversed (i.e., when the perpendicular electric field reverses its polarity). The three dashed horizontal lines (blue, $E^*<0$; black, $E^*=0$; red, $E^*>0$) indicate the values of $E^*$ for which the three $V_{xy}^{2\omega}$ traces in panel \textbf{c} are shown. \textbf{d} Second harmonic longitudinal voltage $V_{xx}^{2\omega}$ measured with an excitation current  $I^{\omega} = 2 \mu$A at frequency $\omega$, as a function of $\delta V_R^B$ and $\delta V_R^T$. $V_{xx}^{2\omega}$ does not change sign upon  reversing the perpendicular electric field, indicating the trivial origin of the nonlinearity in longitudinal resistance. \textbf{e} Calculated amplitude of the longitudinal second harmonic signal $V_{xx}^{2\omega}=\frac{R_{xx}(\delta V_R^B, \delta V_R^T)}{2} (dR_{xx}/d\delta V_R^B + dR_{xx}/d\delta V_R^T)(I^{\omega})^2$ due to  the coupling of the source-drain voltage at frequency $\omega$ to the gate voltages. The calculated signal reproduces the measured one quantitatively, as illustrated in panel \textbf{f} by the comparison of two individual traces, taken along the  vertical dashed black and red lines in panels \textbf{d} and \textbf{e}.}
    \label{fig:FIG2}
\end{figure*} 

\section{Linear and nonlinear transport }
In our experiments, we inject  an alternating current $I^{\omega}$ at a fixed frequency (typically 17.78 Hz),  and measure the longitudinal ($V_{xx}^{\omega}$ and $V_{xx}^{2\omega}$) and transverse voltages ($V_{xy}^{\omega}$ and $V_{xy}^{2\omega}$) at the excitation frequency and at its second harmonic, while sweeping the voltage applied to the top and back gates. From the voltages measured at $\omega$ and at $2\omega$ we infer the linear and nonlinear transport response as a function of electric field and carrier density.

Fig. \ref{fig:FIG1}e shows $R_{xx\square}$ obtained from the longitudinal voltage measured at $\omega$, as a function of the bottom reference potential $\delta V_R^B$, for different values of top reference potential $\delta V_R^T$. The complete evolution of $R_{xx\square}$ is shown in Fig. \ref{fig:FIG1}f. The $R_{xx\square}$ peak position  (corresponding to the charge neutrality point) shifts when varying $\delta V_R^T$ and the peak resistance increases. The peak shift occurs with slightly different slopes upon inverting the sign of both gate voltages, due to the difference in the geometrical capacitances of the two electrolytes for opposite gate polarities (see previous Section). In Fig. \ref{fig:FIG1}g we re-plot the same data as a function of $V^*=(C_T \delta V_R^T -C_B \delta V_R^B)/(C_T + C_B)$ and $E^*=(C_T \delta V_R^T -C_B \delta V_R^B)/(C_T +C_B) $, which determine respectively the accumulated electron density $n$ and the applied perpendicular electric field (see SI for discussion of $n(V^*, E^*)$). Fig. \ref{fig:FIG1}f shows that the peak resistance increases symmetrically with increasing $E^*$ irrespective of polarity, as expected due to the opening of a band gap. 

The transverse and longitudinal voltages measured at $2 \omega$ as a function of $\delta V_R^B$ and $\delta V_R^T$ are shown in Fig. \ref{fig:FIG2}a and \ref{fig:FIG2}d. The behavior of $V_{xy}^{2\omega}$ is better illustrated in Fig. \ref{fig:FIG2}b, where data are plotted as a function of $E^*$ and $V^*$ (selected line cuts of $V_{xy}^{2\omega}$  are shown in Fig. \ref{fig:FIG2}c). A finite nonlinear response is observed both in $V_{xy}^{2\omega}$ and $V_{xx}^{2\omega}$, with a qualitatively different gate voltage dependence in the two cases.  Fig. \ref{fig:FIG2}b and \ref{fig:FIG2}c show unambiguously that $V_{xy}^{2\omega}$ switches sign upon reversing the perpendicular electric field --i.e., it is parity-odd with respect to $E^*$ that controls the sign of inversion breaking-- and that it changes sign as the chemical potential is shifted from the valence to the conduction band (i.e., as $V^*$ crosses zero). This behavior is that expected for a nontrivial NLHE \cite{sodemann.2015,ma.2019,du.2021,ortix.2021,bandyopadhyay.2024,suarez.2025}. Note that the occurrence of the NLHE requires breaking the threefold rotational symmetry in BLG \cite{sodemann.2015,nandy.2019,xiao.2019,du.2021,ortix.2021}. In our devices threefold rotational symmetry is likely broken  by strain caused by the layer employed to electrostatically decouple the two electrolytes (see Fig. \ref{fig:FIG1}a and S1 in SI), which covers nearly the entire substrate and overlaps slightly with the BLG edges, mechanically pinning the bilayer. 

In contrast,  $V_{xx}^{2\omega}$ does not reverses its sign upon reversing $E^*$, implying that the nonlinearity in the longitudinal voltage has trivial origin. The nonlinearity of $V_{xx}^{2\omega}$ can be readily understood, because the current at frequency $\omega$ generates a voltage along the channel ($R_{xx}I^\omega\sin(\omega t)$) that locally modifies the applied gate voltage. As a result, the time-dependent longitudinal voltage  can be written as $V_{xx}(t) = R_{xx}(\delta V_R^B+ R_{xx} I^\omega\sin(\omega t), \delta V_R^T + R_{xx} I^\omega\sin(\omega t))I^\omega\sin(\omega t)$. As a result, $V_{xx}(t)$  contains a $2\omega$ component of amplitude  $V_{xx}^{2\omega}=\frac{R_{xx}(\delta V_R^B, \delta V_R^T)}{2} (dR_{xx}/d\delta V_R^B + dR_{xx}/d\delta V_R^T)(I^{\omega})^2$. We use the measured functional dependence of $R_{xx}(\delta V_R^B, \delta V_R^T)$ to calculate this quantity and find good agreement with the measured gate voltage dependence of  $V_{xx}^{2\omega}$ (compare Fig. \ref{fig:FIG2}c and Fig. \ref{fig:FIG2}d, and black and red curves in Fig. \ref{fig:FIG2}f). This analysis illustrates how the ability to control inversion breaking enables the straightforward delineation of trivial ($V_{xx}^{2\omega}$) and nontrivial ($V_{xy}^{2\omega}$) nonlinear signals, by monitoring their evolution upon $E^*$ parity inversion. 

\begin{figure*}[t]
    \centering
\includegraphics[width=0.7\textwidth]{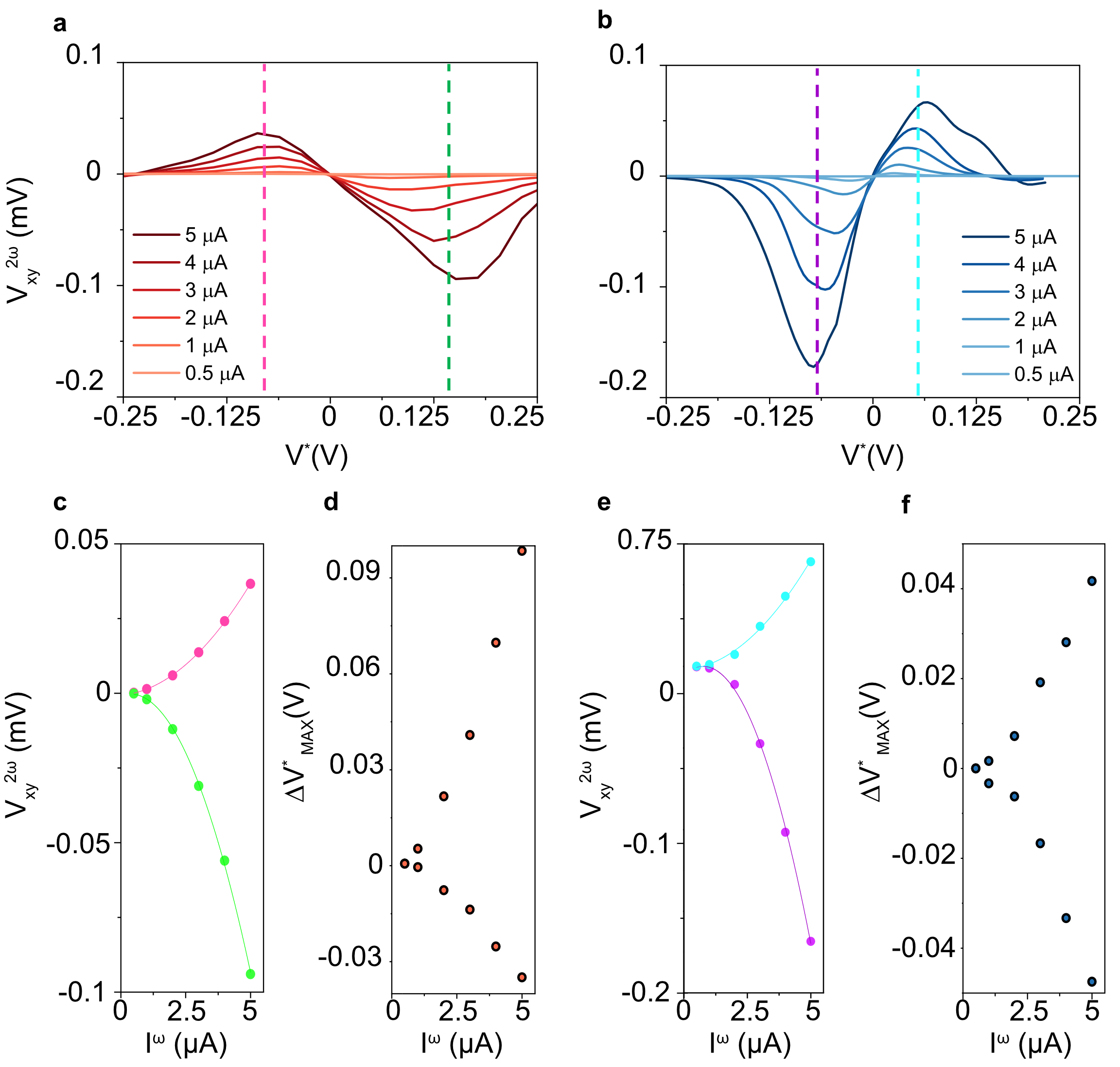}
 \caption{\justifying\textbf{Scaling of second-harmonic signals with excitation current}. \textbf{a} Second harmonic transverse voltage $V_{xy}^{2\omega}$ as a function of $V^*$ for different values of $I^{\omega}$ (see legend), measured for $E^*=0.25V$. \textbf{b} Same measurements as in panel \textbf{a}, taken for the opposite sign of perpendicular electric field, at $E^*= - 0.28V$. \textbf{c}, \textbf{e}  Irrespective of the sign and the value of $V^*$ and of $E^*$, the second-harmonic Hall voltage $V_{xy}^{2\omega}$ scales quadratically with  $I^{\omega}$ (in panel \textbf{c}: $V^* =-0.075V, 0.14V $, $E^*=0.25V$ ; in panel \textbf{e}: $V^* =-0.063V, 0.055V$, $E^*= - 0.28V$ ).  Panels \textbf{a} and \textbf{b} further show that the value of $V^*$ at which $V_{xy}^{2\omega}$ peaks shifts with the applied current $I^{\omega}$. The shift $\Delta V_{MAX}$ increases quadratically with $I^{\omega}$ (see panels \textbf{d} and \textbf{f} for $E*<0$ and $E^*>0$), indicating that the shift is the manifestation of another nonlinear transport phenomenon. }
    \label{fig:FIG3}
\end{figure*}

Fig. \ref{fig:FIG3}a and \ref{fig:FIG3}b illustrate how the amplitude of the NLHE signal $V_{xy}^{2\omega}$ depends on $I^{\omega}$.  For any fixed value of $V^*$, the magnitude of $V_{xy}^{2\omega}$ scales quadratically with $I^{\omega}$, as shown in Fig. \ref{fig:FIG3}c and \ref{fig:FIG3}e, confirming the expected behavior for a second-order nonlinearity. Interestingly, the position of the maximum of $V_{xy}^{2\omega}$ as a function of $V^*$ also shifts with increasing $I^{\omega}$ and the measured shift $\Delta V_{max}$ scales quadratically with $I^{\omega}$ (see Fig. \ref{fig:FIG3}d and \ref{fig:FIG3}f), indicative of another nonlinear effect. Simple heating is unlikely to be at the origin of the effect, since measurements are performed at room temperature and the voltage drop over the entire Hall bar is only a few times the thermal energy $k_BT$ even at the larger current amplitude $I^{\omega}=5 \mu$A. The effect may --at least in part-- arise from a rectified interlayer charge transfer induced by in-plane oscillating electric field \cite{matsyshyn.2023}. Such second-order polarization has a time-independent component that changes the perpendicular electric field between the two layers, and contributes to the perturbation that breaks parity (which is why it can affect the NLHE). More investigations are needed to conclusively determine the origin of the phenomenon.

\begin{figure}[t]
    \centering
\includegraphics[width=0.5\textwidth]{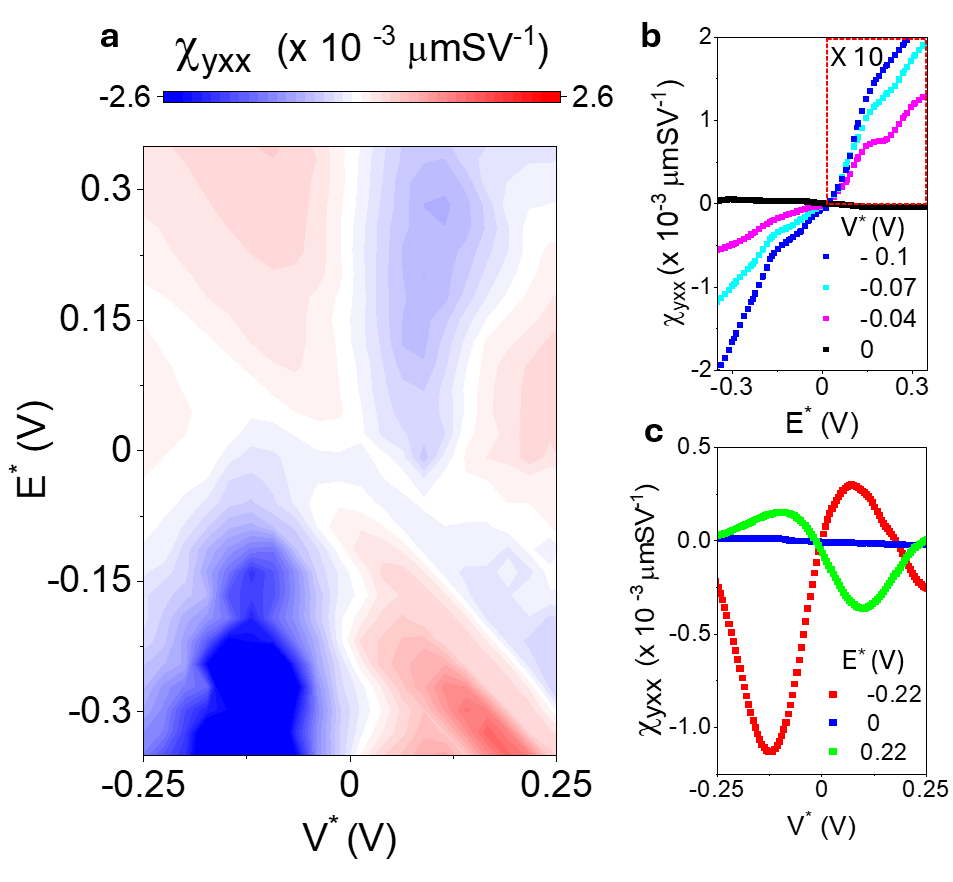}
   \caption{\justifying\textbf{Gate-dependence of the nonlinear susceptibility tensor. a} Component $\chi^{(2)}_{yxx}$ of the nonlinear susceptibility tensor responsible for the nonlinear transverse Hall effect, plotted as a function of $V^*$ and $E^*$. As expected for a nontrivial effect  $\chi^{(2)}_{yxx}$  when the sign of $E^*$ changes (i.e., when the polarity of the perpendicular electric field that breaks parity is reversed) and when $V^*$ changes from negative to positive  (i.e., when the chemical potential is shifted from the valence to the conduction band ). \textbf{b} $\chi^{(2)}_{yxx}$ as a function of $E^*$, for different values of $V^*$ (indicated in the legend). For either polarity of perpendicular electric field,  $\chi^{(2)}_{yxx}$ increases approximately linearly with $E^*$, with absolute magntitude that depends on the sign of $E^*$ (reaching values of 2 10$^{-3}$ $\mu$mS/V for $E^*<0$). \textbf{c} Cuts of the color plot in \textbf{a} for $E^*=\pm 0.22$ V and $E^*=0$.}
    \label{fig:FIG4}
\end{figure}

To properly quantify the strength of the NLHE, we extract the second-order susceptibility tensor $\chi^{(2)}_{yxx} = \dfrac{V_{xy}^{(2\omega)}}{I_x^2}\,\dfrac{\sigma_{xx}^3 W^2}{L^3}$, shown in the color plot of Fig. \ref{fig:FIG4}a. The sign of the second-order response changes when inverting the perpendicular electric field (i.e., when inverting  $E^*$) and when passing through charge neutrality (i.e., when changing the sign of $V^*$). Fig. \ref{fig:FIG4}b further shows that for $E^*<0$  $\chi^{(2)}_{yxx}$ increases approximately linearly with $E^*$ and reaches 2 10$^{-3}$ $\mu$m S/V. This value is comparable to the largest ever reported in any other conductor, despite the fact that our measurements are performed at room temperature. For $E^*>0$, $\chi^{(2)}_{yxx}$ again increases approximately linearly with $E^*$ and reaches values that are still relatively high, but approximately one order of magnitude smaller than in the $E^*<0$ case. 

We follow the same procedure to also estimate the magnitude of the second order longitudinal response, i.e., $\chi^{(2)}_{xxx}$. If we take the measured value of $V_{xx}^{2\omega}$, we find  $\chi^{(2)}_{xxx}$ to be only a few percent of $\chi^{(2)}_{yxx}$. This small value is still a large overestimate because --as we have shown earlier (see Fig. \ref{fig:FIG2})-- $V_{xx}^{2\omega}$ does not change sign upon inverting the perpendicular electric field, and is dominated by the trivial coupling of $V_{xx}^{2\omega}$ to the gate voltages. Estimating the part of the  $V_{xx}^{2\omega}$ signal that is antisymmetric in $E^*$ gives $\chi^{(2)}_{xxx} < 0.001  \chi^{(2)}_{yxx}$, i.e. the non-trivial longitudinal nonlinear response is vanishingly small as compared to the transverse one. These considerations exclude that the NLHE that we observe is due to the valley-contrasting chirality effect recently reported in graphene-on-hBN devices, for which the longitudinal and transverse nonlinear response have comparable magnitudes \cite{isobe.2020,he.2022}.

Finally, we briefly discuss the microscopic origin of the observed NLHE. As our devices are operated at room temperature, with ions in the electrolytes in direct contact with the BLG causing modest values of carrier mobility ($\approx 1000-2000$ cm$^2$/Vs), the dominant mechanism responsible for the NLHE should be expected to scale with the lowest possible power of the scattering time $\tau$. Within a semiclassical treatment, the lowest power of $\tau$ is the one expected if the NLHE is due to a Berry curvature dipole (BCD; the contribution is linear in $\tau$)\cite{sodemann.2015,matsyshyn.2019,du.2019,nandy.2019,xiao.2019,du.2021b,ortix.2021,suarez.2025}. However, as it can be established by comparing the symmetric dependence of the linear resistance $R_{xx\square}$ with perpendicular electric field, to the pronounced  asymmetry in the magnitude of $\chi^{(2)}_{yxx}$ upon reversing $E^*$, our data is not compatible with the BCD giving the dominating contribution. Indeed, Fig. \ref{fig:FIG1}e-g show that  the peak in $R_{xx\square}$ increases in an almost perfectly symmetric way, irrespective of the sign of $E^*$, as expected because the magnitude of band gap that opens in BLG does not depend on the sign of $E^*$. If the BCD contribution to  the NLHE --which also  depends on $E^*$ through the bandgap-- was dominant, the magnitude of  $\chi^{(2)}_{yxx}$ should therefore also be symmetric in $E^*$. The asymmetry in  $\chi^{(2)}_{yxx}$ upon reversing $E^*$ therefore implies that --while a BCD contribution may be present-- it is not the dominant mechanism responsible for the observed NLHE. 

We believe that the dominant mechanism originates from asymmetries in the scattering of charge carriers, i.e. the so-called skew scattering and side-jump terms. Such a mechanism naturally explains why the magnitude of the NLHE differs significantly when reversing the polarity of the perpendicular electric field, because the type of ions in contact with the BLG in the top and bottom electrolytes changes when the polarity of $E^*$ is reversed, and so is the microscopic nature of the asymmetry in scattering potential. While in the semiclassical approximation collision-induced NLHE is predicted to scale with powers larger-than-linear in $\tau$ \cite{du.2019,bandyopadhyay.2024}, NLHE theories  that go beyond the semiclassical approximation \cite{du.2021b}, find  collision-induced contributions terms that contribute to the NLHE and are predicted to  scale linearly with $\tau$ \cite{nandy.2019}(i.e., with the same power of $\tau$ as in the BCD mechanism). Our data appear to be consistent with these theoretically predicted contributions.

\section{Discussion}
The magnitude of the room-temperature NLHE in double ionic gated Bernal-stacked graphene is comparable to that of the effect observed in  WTe$_2$ or in graphene-based moiré superlattice devices  at cryogenic temperatures (4 K or below) \cite{zhang.2018,ma.2019,kang.2019,wang.2019,zeng.2019,tiwari.2021,he.2022,sinha.2022,huang.2023}, and much larger than values reported at room temperature in TaIrTe$_4$, BaMnSb$_2$, or in Bi \cite{kumar.2021, min.2023, makushko.2024}. To compare to existing measurements at room temperature, we look at the  nonlinear Hall voltage measured at a given current, and consider that the signal scales with the current squared. At room temperature, TaIrTe$_4$ devices show a  nonlinear Hall voltage of 100  $\mu$V --i.e., a signal comparable to the one we measure (see Fig. \ref{fig:FIG3})-- when the  current flowing is $I = 600 \mu$A \cite{kumar.2021}, whereas in our devices the current needed to achieve the same voltage is only $5 \mu$A. Considering the scaling with $I^2$, the magnitude of the NLHE in double ionic gated bilayer devices is a factor of $10^3-10^4$ larger than in TaIrTe$_4$. In BaMnSb$_2$ devices, the nonlinear Hall voltage measured at room temperature is approximately 50 $\mu$V for a current $I= 100 \mu$A, corresponding to a signal $10^3$ times smaller than in our devices (for Bi devices, the signal is yet much smaller) \cite{min.2023, makushko.2024}. \\

These comparisons highlight the truly giant magnitude of the room-temperature NLHE  in double ionic gated bilayer graphene. Because the NLHE provides a zero‑bias rectification mechanism – without the need to overcome a built‑in potential as in conventional diode rectifiers – a large nonlinear Hall response directly translates into an efficient conversion of an AC field into a DC signal. The efficiency of this rectification mechanism can be quantified by the voltage responsivity \cite{isobe.2020, hemour.2014}, defined as the ratio between the generated DC voltage and power dissipation:
\[
R_V = \frac{1}{L}\,\frac{V_{xy}^{0}}{\sigma_{xx} E_x^2}
    = \frac{1}{L}\,\frac{\chi_{yxx}^{(2)}}{\sigma_{xx}^2},
\]
where $V_{xy}^{0}$ is the rectified transverse voltage. An efficient NLHE rectifier must therefore combine a large $\chi_{yxx}^{(2)}$ with a low $\sigma_{xx}$ (i.e., a large $R_{xx\square}$), two conditions simultaneously fulfilled in our devices. From the measured values of $\chi_{yxx}^{(2)}$ (see Fig.~\ref{fig:FIG4}a) and of $R_{xx\square}$; see Fig.~\ref{fig:FIG1}e) we obtain a room‑temperature voltage responsivity $R_V \simeq 4\times 10^{4}\,\mathrm{V/W}$ near charge neutrality at the largest applied perpendicular electric field.\\

This $R_V$ value places double ionic gated BLG among the best rectifiers reported to date, with a voltage responsivity comparable to that of currently available commercial Schottky diodes ($\sim 1\times 10^{5}$ – $5\times 10^{4}\,\mathrm{V/W}$) \cite{hemour.2014}. Previously studied systems, including Weyl semimetal TaIrTe$_4$ , KTaO$_3$‑based two‑dimensional electron gases , and the Dirac material BaMnSb$_2$\, exhibit room‑temperature voltage responsivities of order $0.1\,\mathrm{V/W}$ \cite{kumar.2021}, $0.2\,\mathrm{V/W}$ \cite{zhai.2023}, and $101\,\mathrm{V/W}$ \cite{min.2023}, respectively, all several orders of magnitude smaller than in our case. Double ionic gated BLG therefore provides a 2D, electrostatically tunable platform that combines a giant NLHE with ultrahigh room‑temperature voltage responsivity, offering true  potential for practical high-frequency rectification and detection, potentially deep in the THz regime.\\

\section*{Acknowledgments}
We gratefully acknowledge A. Ferreira for technical support, E. Linardy and C. Cao for contributions to the development of the technique of double ionic gating, G.Sala for helpful discussions on the nonlinear Hall measurements and their interpretation. AFM acknowledges financial support from the Swiss National Science Foundation, Division II, under  grant 200021-227636. J.S. acknowledges support from Singapore Ministry of Education (MOE) Academic Research Fund Tier 3 Grant No. (MOE-MOET32023-0003) “Quantum Geometric Advantage”.
%%%%%%%%%%%%%%%%%%%%%%%%%%%%%%%%%%%%%%%%
\clearpage
\onecolumngrid
\setcounter{section}{0}
\renewcommand{\thesection}{\arabic{section}}
% From here onward, write entries into the TOC again
\let\addcontentsline\savedaddcontentsline

\title{Supplementary Material  \texorpdfstring{\\}{}``Switchable giant room-temperature nonlinear Hall effect in Bilayer Graphene"}

\author{Margherita Melegari}
\email{margherita.melegari@unige.ch}
\affiliation{Department of Quantum Matter Physics, University of Geneva, Quai Ernest-Ansermet 24, 1211 Geneva, Switzerland}
\affiliation{Department of Applied Physics, University of Geneva, 24 Quai Ernest Ansermet, Geneva, CH-1211 Switzerland}
\author{Ignacio Gutierrez Lezama}
\affiliation{Department of Quantum Matter Physics, University of Geneva, Quai Ernest-Ansermet 24, 1211 Geneva, Switzerland}
\affiliation{Department of Applied Physics, University of Geneva, 24 Quai Ernest Ansermet, Geneva, CH-1211 Switzerland}

\author{Justin C. W. Song}
\affiliation{Division of Physics and Applied Physics, School of Physical and Mathematical Sciences, Nanyang Technological University, Singapore 637371, Singapore}
\author{Alberto F. Morpurgo}
\email{alberto.morpurgo@unige.ch}
\affiliation{Department of Quantum Matter Physics, University of Geneva, Quai Ernest-Ansermet 24, 1211 Geneva, Switzerland}
\affiliation{Department of Applied Physics, University of Geneva, 24 Quai Ernest Ansermet, Geneva, CH-1211 Switzerland}
\maketitle

{\tableofcontents\par}
\clearpage
\section{Device fabrication and characterization} To realize the BLG double ionic gated devices employed in this work we exfoliate BLG on a Si/SiO$_2$ substrate, transfer it with a polymeric stamp onto a LICGC substrate, and attach electrodes by means of conventional nanofabrication techniques (electron-beam lithography, Pt/Au evaporation, and lift-off; as discussed in Ref. \cite{cao.2023}, device stability requires that direct contact between metal and LICGC is avoided, which we achieve by  electron-beam evaporating 40 nm of SiO$_2$ between the metal electrodes and the substrate). A SiO$_2$ layer (100 nm) covered by an Al/Al$_2$O$_3$ layer (20nm), and a PMMA layer are deposited overlapping with the edges of the BLG (approximately 100 nm), leaving uncovered the majority of the Hall bar, to decouple electrostatically the top and bottom electrolytes and prevent cross-talk between the gate voltages \cite{domaretskiy.2022, melegari.2026}. The top gate and reference electrodes consist of very large Cr/Au pads evaporated onto the PMMA through a shadow mask. The bottom gate electrode is an evaporated Cr/Au layer covering the backside of the substrate; the bottom reference potential is a nanostructured Pt/Au contact on top of the LICGC substrate, a few tens of micron away from the BLG. 

The ionic liquid is applied on top of the substrate immediately prior to loading the devices in the vacuum chamber where measurements are done, covering the BLG, the gate and the reference electrodes on top of the PMMA layer. Electrical measurements are performed in  a cryofree Teslatron Oxford Instrument cryostat, operated with the sample near room temperature to ensure that ions in the electrolyte are mobile. The cryostat is equipped with a superconducting magnet to apply magnetic field and measure the linear Hall effect, as one of the characterization steps of our devices. Measurements are performed with the devices in vacuum (p $<10^{-6}$ mbar). 

The top and back gate voltages are initially swept gradually (and slowly, few mV/s) over an increasingly large range, reaching a maximum value of $\pm 2.5$ V, a procedure used to remove adsorbates from the gated surface. The current through the two gates electrodes is monitored at all times, to ensure that it remains below 1 nA (the current measured is the displacement current due to charging of the capacitors; if the gate voltages are keep constant, no current is measured within the sensitivity of our system, approximately 0.1 nA). The absence of leakage current is important because in ionic liquid gated system leakage current exclusively results from chemical reactions between the electrolytes and the gated material, which cause device degradation. The absence of chemical reactions is confirmed by the fact that our devices are stable for a period of months, without significant changes in their electrical characteristics. 

All transport data are analyzed as a function of the potential $\delta V_R^B$ and $\delta V_R^T$ measured at the reference electrodes (which physically correspond to the voltages dropping across the bottom and top gate capacitances $C_B$ and $C_T$), to eliminate the effects of built-in field and of bias stress effects. To be precise, $\delta V_R^B$ and $\delta V_R^T$  are the potentials measured at the reference electrodes relative to the same potentials measured at charge neutrality under zero perpendicular electric field (i.e., the values measured when the device is biased at the peak of the longitudinal resistance, with the height of the peak that is minimum, i.e., red curve in Fig. 1e in the main text). 

More information about the behavior of different types of --single and double-- ionic gate devices based on LICGC substrates can be found in our earlier work, which also discusses in detail all characterization measurements that we have done on different experimental systems (transition metal dichalchogenides and bilayer graphene) to ensure the absence of artifacts affecting the experimental results \cite{domaretskiy.2022,cao.2023,melegari.2026}. 
\newpage
\section{Estimate of the perpendicular electric field at charge neutrality} We estimate the electric field $E_{BLG}$ present between the two graphene layers as a function of $\delta V_R^B$ and $\delta V_R^T$. We use classical electrostatics to determine the applied electric field,  from which we subtract the screening electric field $ \delta ne/\epsilon \epsilon_0$ due to the density of electrons $\delta n$ displaced from one layer to the other ( here $\epsilon \simeq 3$ is the dielectric constant of BLG due to the electrons forming the $\sigma$ bonds \cite{slizovskiy.2021}, and $\epsilon_0$ is the vacuum dielectric constant). In the regime considered here --with the interlayer potential energy difference smaller or comparable to the interlayer hopping integral-- a simple tight-binding model calculation gives $\delta n =\alpha E_{BLG}$ (with $\alpha = 1.34 \ 10^8$ V$^{-1}$ $m^{-1}$), from which we obtain:

\begin{equation}
\begin{split}
E_{BLG} &=
\frac{C_T\delta V_R^T - C_B\delta V_R^B}
{2\epsilon \epsilon_0 \left(1+e\alpha/\epsilon \epsilon_0 + d (C_T+C_B)/4 \epsilon \epsilon_0\right)}
\\
&=
\frac{(C_T+C_B)E^*}
{2\epsilon\epsilon_0 \left(1+e\alpha/\epsilon \epsilon_0 + d (C_T+C_B)/4 \epsilon \epsilon_0\right)} ,
\end{split}
\label{eq:EBLG}
\end{equation}
(in the last step we have written $E_{BLG}$ as a function of the quantity $E^*$ used to plot the data in the main text). Note that in conventional double gated BLG devices, where the gate dielectrics are much thicker than the BLG, the last term in the denominator is negligible and is normally omitted. In double ionic gated devices where the thickness of the ionic layers are approximately 0.1 nm (i.e. thinner than the BLG), considering this last term is essential to have a correct quantitative estimate of the perpendicular electric field. For the largest value of $E^* = 0.35$ V for which the NLHE has been shown in the main text, the
electric field between the graphene layer is $E_{BLG} \simeq 1.7-1.8$ V/nm, giving a value of $\Delta \simeq 0.6-0.7$ eV, somewhat larger than  the interlayer hopping integral $t_{\perp}\approx 0.4$ eV.
\newpage
  \section{Device Stability and Reproducibility}
     In contrast to double gated bilayer graphene (BLG) devices equipped with conventional gates, which have been studied for almost two decades, it is only recently that ionic gates have been used to double gate atomically-thin carbon layers. Because ionic gating is less commonly employed (and known) in the scientific community as compared to ordinary gating with solid-state dielectrics,  here we show additional measurements to illustrate the reproducibility and stability of double ionic gated devices, and to show that their behavior conform to that expected from conventional double gated devices, in the gate voltage range discussed in this paper. 
     
\begin{figure}[h!]
    \centering
    \includegraphics[width=1\linewidth]{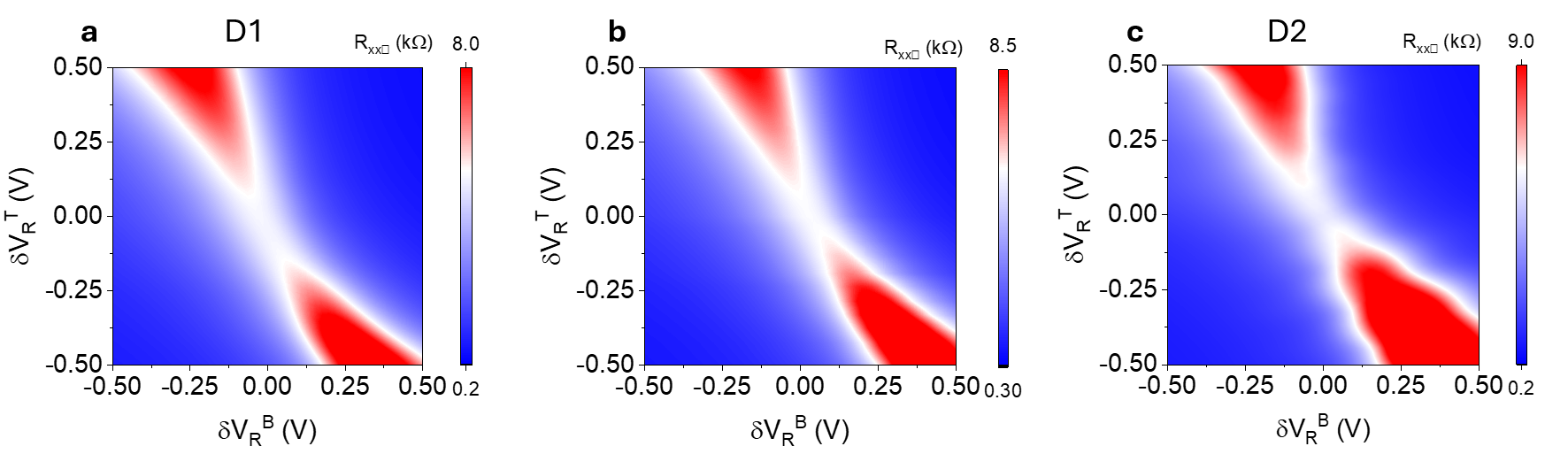}
    \caption{\justifying\small \textbf{Evolution of $R_{xx\square}$ upon double ionic gating BLG.}
    \textbf{(a--b)} Color maps of $R_{xx\square}$ vs $\delta V_R^{\mathrm{T}}$ and $\delta V_R^{\mathrm{B}}$, measured in device D1 using different pairs of voltage probes and \textbf{(c)} $R_{xx\square}$ measured in device D2.}
    \label{fig:S1}
\end{figure}

     Fig. \ref{fig:S1} shows additional data from device D1 (whose data are shown in the main text) and D2 (a second nominally identical device) obtained from measurements done with different pairs of voltage probes, to show the reproducibility of our devices. As can be seen in Fig.\ref{fig:S1}, the color maps of $R_{xx\square}$ Vs $\delta V_R^B$ and $\delta V_R^T$ show a very similar quantitative and qualitative behavior.

\newpage

\begin{figure}[h!]
    \centering
    \includegraphics[width=0.8\linewidth]{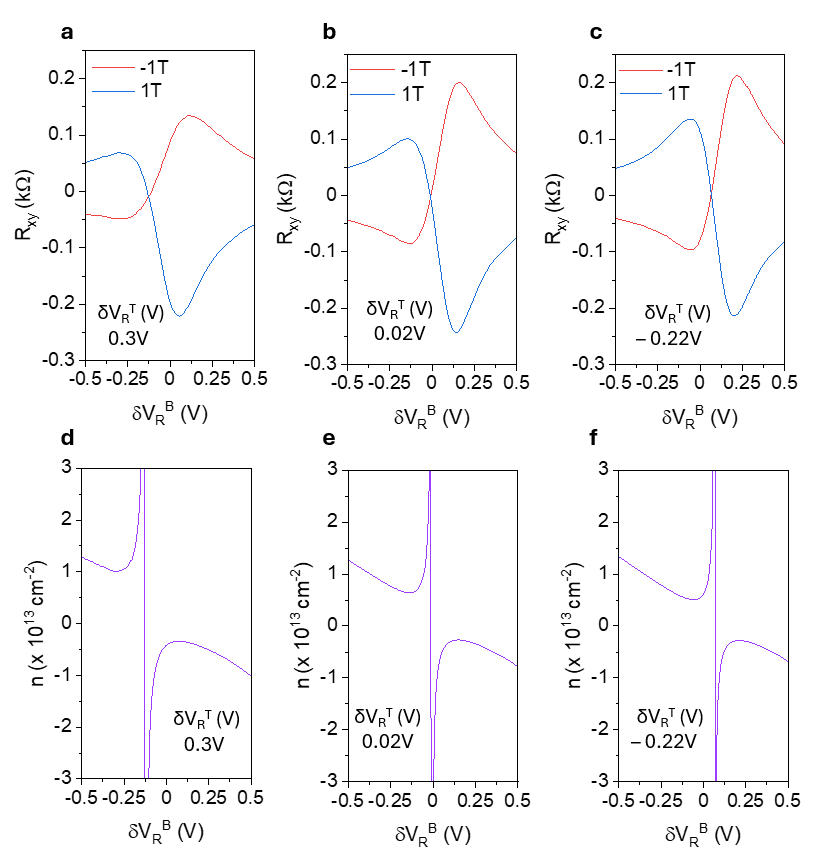}
 \caption{\textbf{Evolution of $R_{xy}$ and $n$ in a double ionic gated device.}
$R_{xy}( \delta V_R^B)$ measured in device D1 at $B = -1\,\mathrm{T}$ (red) and $1\,\mathrm{T}$ (blue) for $\delta V_R^T = 0.3\,\mathrm{V}$ \textbf{(a)}, $0.02\,\mathrm{V}$ \textbf{(b)} and $-0.22\,\mathrm{V}$ \textbf{(c)}. By taking the difference of these signal and dividing by the differences of applied magnetic field we extract the carrier density $n$, whose values are shown in \textbf{(d-f)} for $\delta V_R^T = 0.3\,\mathrm{V}$, $0.02\,\mathrm{V}$ and $-0.22\,\mathrm{V}$.}
    \label{fig:S2}
\end{figure}

  To further illustrate that ionic double gated devices function as expected, in Fig. \ref{fig:S2} we also show the evolution of the transverse resistance (linear Hall resistance) $R_{xy}$ and the corresponding carrier density as a function of $\delta V_R^B$ for selected values of  $\delta V_R^T$ upon double gating of BLG. As expected, $R_{xy}$ changes sign at the same values of $V_R^{\mathrm{B}}$ and $V_R^{\mathrm{T}}$ at which the $R_{xx\square}$ peak is observed. This can be seen by a direct comparison of Fig. \ref{fig:S1}, which shows $R_{xx\square}$ Vs $\delta V_R^{\mathrm{T}}$ and $\delta V_R^{\mathrm{B}}$, and Figs. \ref{fig:S2} (a-c), which show $R_{xy}$ (measured at $B = \pm 1\,\mathrm{T}$) as a function of $\delta V_R^{\mathrm{B}}$ for different values of $\delta V_R^{\mathrm{T}}$ in device D1. In both cases charge neutrality shifts towards more positive values of $\delta V_R^{\mathrm{B}}$ as $\delta V_R^{\mathrm{T}}$ decreases.

  The charge carrier density calculated from the $R_{xy}$ data according the relation $1/en = R_{H} = [R_{xy}(\delta V_R^{\mathrm{B}},1T)-R_{xy}(\delta V_R^{\mathrm{B}},-1T)]/\Delta B$ is shown in Figs. \ref{fig:S2} (a-c),  The extracted densities are shown in Fig. \ref{fig:S2} (d-f). Away from charge neutrality the carrier density evolves linearly with $\delta V_R^{\mathrm{B}}$, as expected for the electrostatic accumulation of charge carriers. In contrast to the value of the electric field at charge neutrality, which is determined by the geometrical capacitances of the two electrolyte gates, the value of density away from charge neutrality  is determined by the quantum capacitance of BLG (which for ionic liquid gated devices is much smaller than the geometric capacitances, $\approx 50 \mu$F/cm$^2$).
 
\clearpage
    \section{Relation between $V^*$ and $n$}
\begin{figure}[h!]
    \centering
    \includegraphics[width=1\linewidth]{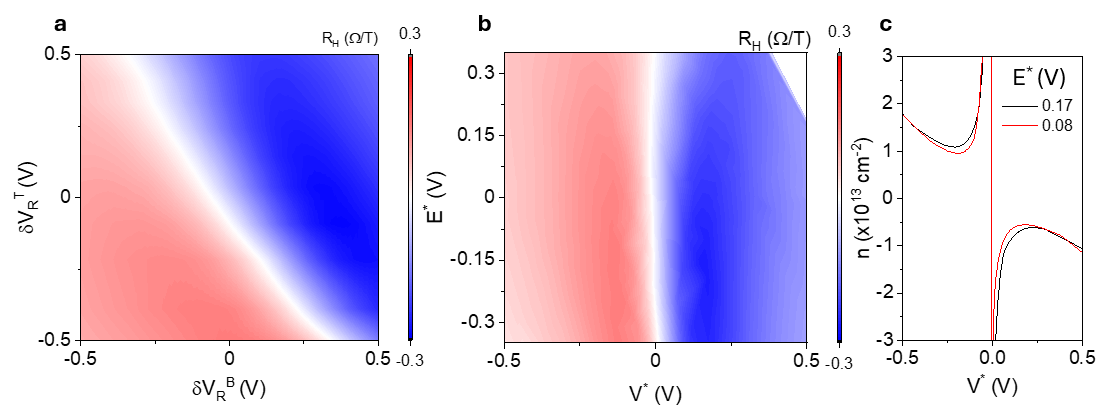}
    \caption{\textbf{Evolution of $n$ as a function of $V^*$.}
    \textbf{(a)} Color plot of $R_H$ as a function of $\delta V_R^B$ and $\delta V_R^T$ and \textbf{(b)} as a function of $V^*$ and $E^*$ extracted from measurements of $R_{xy}$ performed on device D1. \textbf{(c)}. Dependence of the carrier density $n$ (extracted from $R_H$ in \textbf{(b)}) on $V^*$, for two different values of $E^*$ (see legend).}
    \label{fig:S3}
\end{figure}

    In the main text we mention that $V^*$ and $E^*$ (calculated from $\delta V_R^{\mathrm{B}}$ and $\delta V_R^{\mathrm{T}}$; see main text) are determined respectively by the electron density $n$ and the perpendicular electric field across BLG $E_{BLG}$. Here we show that $n$ does indeed depend on $V^*$ and is independent of $E^*$. To this end, we have extracted the Hall coefficient $R_{H}$ from the $R_{xy}$ measured as a function of $\delta V_R^{\mathrm{B}}$ and $\delta V_R^{\mathrm{T}}$ in device D1 (see Fig. \ref{fig:S3} a),  and plotted it as a function of $V^*$ and $E^*$ in Fig. \ref{fig:S3} b. $R_{H}$ indeed evolves when we vary $V^*$ but remains nearly  perfectly constant when we vary $E^*$.\\
    
    That $n$ is primarily controlled by $V^*$ and is largely independent of $E^*$ within the explored range can also be explicitly seen in Fig. \ref{fig:S3} c, which shows $n(V^*)$ for different values of $E^*$. Note in passing that the values of $n$ away from charge neutrality ($V^*=0)$ are determined by the quantum capacitance of BLG and not by the geometrical capacitance of the ionic gates, which are significantly larger and only play a role at charge neutrality. We emphasize this point because the ratio of $C_B/C_T$ extracted from the evolution of the resistance peak (charge neutrality) with $\delta V_R^{\mathrm{B}}$ as $\delta V_R^{\mathrm{T}}$ (see main text) --which we use to compute the values of $V^*$ and $E^*$-- corresponds to that of the geometrical capacitance of the ionic gates.

 \newpage
  \section{Spurious second-order ${\displaystyle \chi_{xxx}}$ response}
\begin{figure}[h!]
    \centering
    \includegraphics[width=0.4\linewidth]{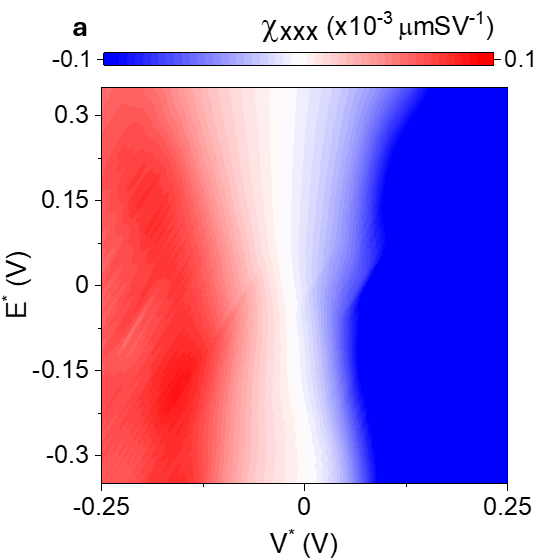}
    \caption{\textbf{Color plot of $\chi^{(2)}_{xxx}$ as a function of $V^*$ and $E^*$.}
    Plot of $\chi^{(2)}_{xxx}$ obtained from the second-harmonic longitudinal signal using the same expression employed to calculate $\chi^{(2)}_{yxx}$ (i.e. $ \chi^{(2)}_{xxx} = \dfrac{V_{xx}^{(2\omega)}}{R_{xx}^3}\,\dfrac{L}{I_x^2}$. the magnitude of  $\chi^{(2)}_{xxx}$ calculated in this way is only a few percent of $\chi^{(2)}_{yxx}$. Note that --as expected for a trivial nonlinearity-- $\chi^{(2)}_{xxx}$  is symmetric under inversion of $E^*$.}
    \label{fig:S4}
\end{figure}

  In the main text we show that the non-linear longitudinal voltage $V_{xx}^{2\omega}$ is a spurious effect originating by the coupling of the applied gate voltage to the AC signal sent through the channel of the device to measure the non-linear Hall voltage $V_{xy}^{2\omega}$. Nevertheless, to determine a bound to possible nonlinear effects in the longitudinal resistance, it is useful to use the measured $V_{xx}^{2\omega}$  data and calculate the corresponding value of $\chi^{(2)}_{xxx} = \dfrac{V_{xx}^{(2\omega)}}{R_{xx}^3}\,\dfrac{L}{I_x^2}$. As can be seen in the color plot in Fig. \ref{fig:S4}, the order of magnitude of $\chi^{(2)}_{xxx}$ --which is symmetric with respect to $E^*$-- is 30 times smaller than those of $\chi^{(2)}_{yxx}$ (i.e., $\chi^{(2)}_{xxx}$ is a few percent of $\chi^{(2)}_{yxx}$; for comparison see Fig.4). Since any nontrivial nonlinear effect has to change sign upon inverting the sign of $E^*$, we have also looked at the antisymmetric component of $\chi^{(2)}_{xxx}$ with respect to $E^*$ and found that its magnitude if of the order of $10^{-6}\mu$ mSV$^{-1}$, i.e., 3 orders of magnitude smaller than $\chi^{(2)}_{yxx}$. In our devices, therefore we find no experimental evidence for the presence of a non-trivial non-linear longitudinal signal within our experimental sensitivity.

\newpage

 \section{Reproducibility of the non-linear response}
\begin{figure}[h!]
         \centering
         \includegraphics[width=0.9\linewidth]{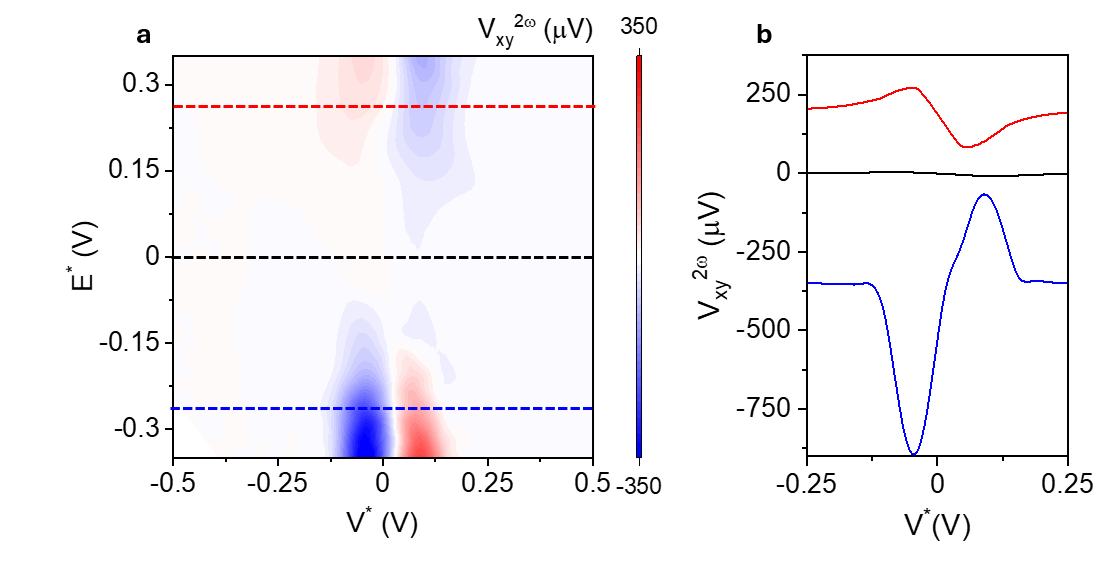}
         \caption{\textbf{Evolution of $V_{xy}^{2\omega}$ measured in device D2 as a function of $V^*$ and $E^*$.} 
        \textbf{a} Color plot of $V_{xy}^{2\omega}$ measured at $I^{\omega} = 2\,\mu\text{A}$ and $\omega = 17.78 Hz$, plotted as a function of $V^*$ and $E^*$. The three horizontal dashed lines indicate the representative values of $E^*$ (blue for $E^*<0$, black for $E^*=0$ and red for $E^*>0$) at which line-cuts of $V_{xy}^{2\omega}$ Vs $V^*$ plotted in \textbf{(b)} were taken.}
    \label{fig:S5}
     \end{figure}

 To demonstrate the reproducibility of the non-linear response present in our devices, in this section we show additional data measured on device D2, showing that the behavior of $V_{xy}^{2\omega}$ is entirely consistent with the that observed in device D1. Fig. \ref{fig:S5} shows that $V_{xy}^{2\omega}$ measured in device D2 (at an excitation current $I^{\omega} = 2\,\mu\text{A}$ at frequency $\omega$) changes sign upon inversion of $V^*$, and upon inversion of $E^*$, reflecting its odd parity nature (see Fig. 2b and c). This is the exact same behavior shown in Fig. 2b and c for device D1 (panel b) shows cuts of panel a) taken along the dashed lines of the corresponding color; in panel c), curves are offset for clarity).

     \newpage
    
\begin{figure}[h!]
    \centering
    \includegraphics[width=0.9\linewidth]{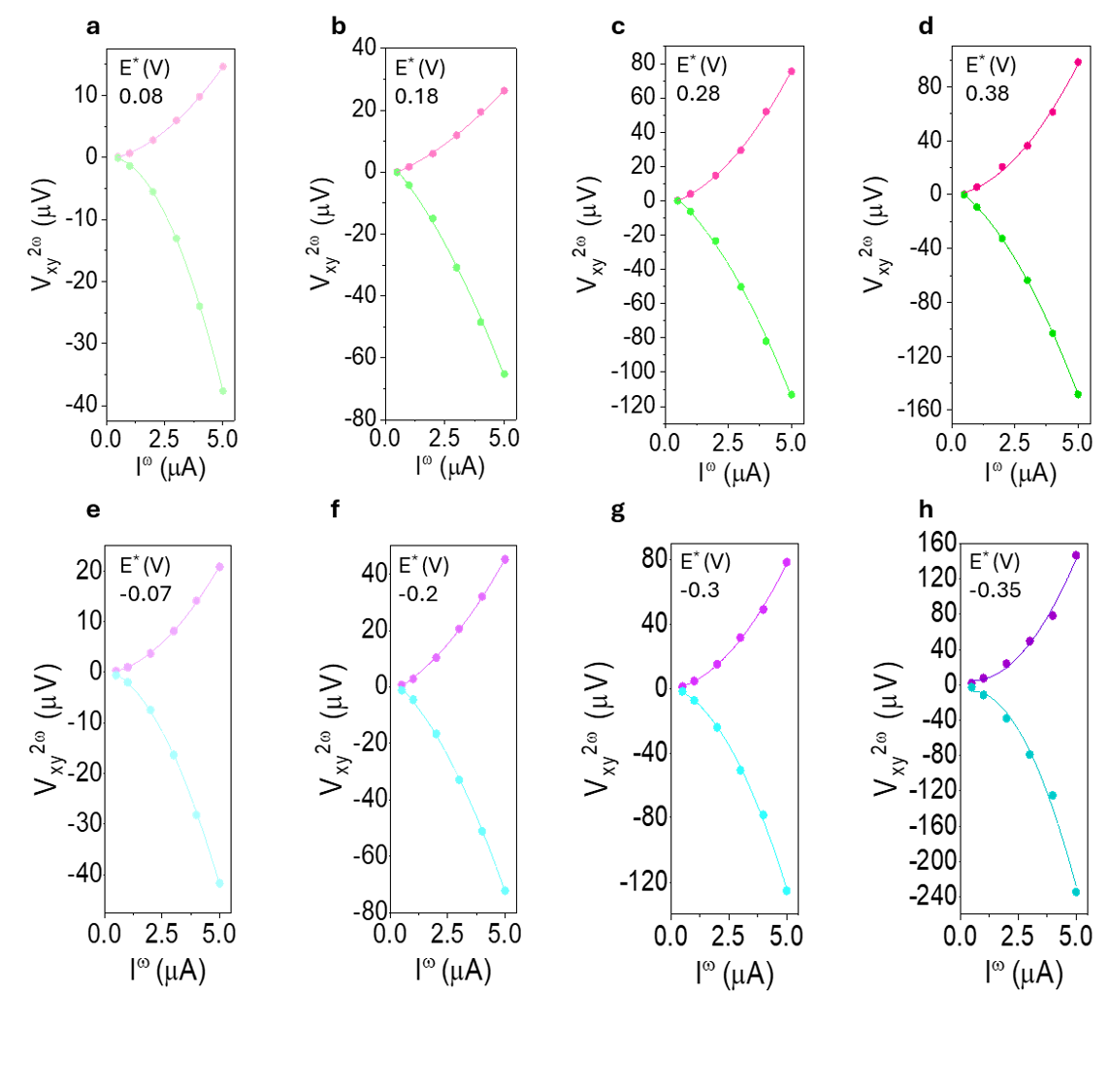}
\caption{\textbf{Quadratic scaling of of $V_{xy}^{2\omega}$ with $I^{\omega}$ and its evolution with $E^*$.}
\textbf{a–h} Second-harmonic transverse voltage $V_{xy}^{2\omega}$ measured on device D2 at fixed $V^*$ plotted as a function of $I^{\omega}$ for different values of $E^*$. \textbf{a–d} correspond to $E^* = 0.08$ V, $0.19$ V, $0.28$ V, and $0.38$ V, respectively, while \textbf{e–h} correspond to $E^* = -0.07$ V, $-0.20$ V, $-0.30$ V, and $-0.35$ V, respectively.} 
  \label{fig:S6}
\end{figure}

  Fig. \ref{fig:S6} shows the dependence of $V_{xy}^{2\omega}$ on $I^{\omega}$ measured on Device D2 for  increasingly large positive (Fig.~\ref{fig:S6} (a–d)) and negative (Fig. \ref{fig:S6} (e–h)) values of $E^*$.  $V_{xy}^{2\omega}$ increases quadratically with $I^{\omega}$ and its magnitude for a given value of $I^{\omega}$ increases systematically with $|E^*|$, which is precisely  the behavior observed for device D1 and discussed in the main text.

\bibliographystyle{myapsrev}
% Bibliography
\newpage
\input{References.bbl}
\end{document}